\documentclass[showpacs,twocolumn,floats]{revtex4}
\usepackage{graphicx}
\usepackage{bm}
\begin{document}
\title{Breakdown of Universality in Quantum Chaotic Transport:\\
the Two-Phase Dynamical Fluid Model}
\author{Ph.~Jacquod and E.V. Sukhorukov}
\affiliation{D\'epartement de Physique Th\'eorique,
Universit\'e de Gen\`eve, CH-1211 Gen\`eve 4, Switzerland} 
\date{November 22, 2003}
\begin{abstract}
We investigate the transport properties of open quantum chaotic systems in 
the semiclassical limit. We show how the transmission spectrum, the
conductance fluctuations, and their correlations are influenced by the
underlying chaotic classical dynamics, and result from the separation of
the quantum phase space into a stochastic and a deterministic phase.
Consequently, sample-to-sample conductance fluctuations lose their 
universality, while the persistence of
a finite stochastic phase protects the universality  of
conductance fluctuations under variation of a quantum parameter.
\end{abstract}
\maketitle
Universal Conductance Fluctuations (UCF) 
are arguably one of the most spectacular manifestations
of quantum coherence in mesoscopic systems \cite{ucf}. In metallic
samples, the universality of the conductance fluctuations manifests 
itself in their magnitude, ${\rm rms} (g) = O(e^2/h)$, 
independently on the sample's shape and size, 
its average conductance or the exact configuration of the underlying
impurity disorder \cite{ucf,webb}. 
In ballistic chaotic systems, a similar behavior is
observed, which is captured by Random Matrix Theory (RMT)
\cite{ucfrmt}. For an open chaotic cavity connected
to two $N$-channel leads, and thus having an average classical conductance 
$g=N/2$ (we consider spinless fermions and
express $g$ in units of $e^2/h$), 
RMT predicts a universal conductance variance $\sigma^2(g)=1/8$
for time-reversal and spin rotational symmetric
samples. At the core of the UCF lies the {\it ergodic hypothesis} that
sample-to-sample fluctuations are equivalent to fluctuations induced
by parametric variations (e.g. changing the
energy or the magnetic field) within a given 
sample \cite{ucf}.

According to the scattering theory, 
transport properties such as the conductance 
derive from transmission eigenvalues, i.e.  $g = \sum_{i=1}^N T_i$ 
\cite{buttiker}. 
While coherence effects such as the UCF arise due to nontrivial
correlations $\langle T_i T_j \rangle$, the knowledge of the probability
distribution $P(T)$ of
transmission eigenvalues is sufficient to correctly
predict, e.g., the average conductance, or 
the Fano factor $F \equiv \langle T(1-T) \rangle / \langle T\rangle$ 
for the shot-noise power \cite{blanter}.
For a ballistic chaotic cavity, RMT predicts \cite{ucfrmt}
\begin{equation}\label{probt}
P_{\rm RMT}(T) = \frac{1}{\pi} \frac{1}{\sqrt{T(1-T)}},
\end{equation}
and thus $F=1/4$. For shot-noise, as well as for UCF, the correct universal
behavior is captured by RMT.

\begin{figure}
\includegraphics[width=5.6cm,angle=270]{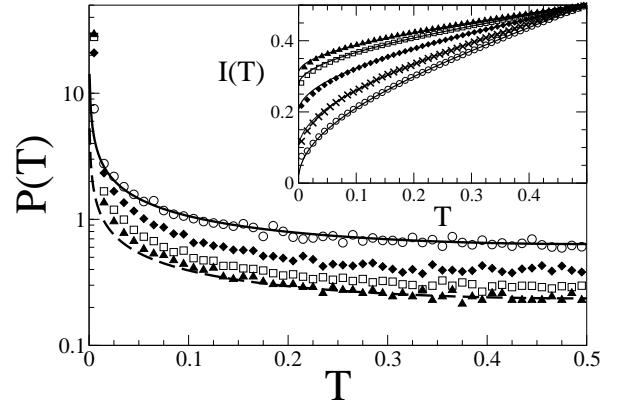}
\caption{
Distribution of transmission eigenvalues for $K=27.65$, 
$\tau_D=25$, and $M=2048$ (empty circles;
$\tau_E \simeq 0$; distribution calculated over 729 different samples); 
$K=9.65$, $\tau_D=5$, and 
$M=1024$ (black diamonds; $\tau_E=1.5$; 729 samples); 
$M=8192$ (empty squares; $\tau_E=2.8$;
16 samples); and $M=65536$ 
(black triangles; $\tau_E=4.1$; 1 sample). The solid line
gives the universal distribution $P_{\rm RMT}$ of Eq.~(\ref{probt}),
and the dashed line the distribution $P_{\alpha}$ 
of Eq.~(\ref{probtalpha}), with $\alpha = 0.39$. Note that
$P(T)$ is symmetric around $T=0.5$.
Inset: integrated probability distribution of transmission
eigenvalues for the same set of parameters as in the main panel,
as well as for $K=9.65$, $\tau_D=5$, and 
$M=128$ ($\times$; $\tau_E=0.16$).
The solid lines are fits obtained from Eq.~(\ref{intprobtalpha}), with
$\alpha \approx 0.98$, 0.81, 0.6, 0.45 and 0.385 (from bottom to top).}
\label{figure4}
\end{figure}

The validity of RMT is however generically 
restricted by the existence of finite
time scales. Spectral fluctuations are known to 
deviate from RMT predictions for energies larger than the inverse
period of the shortest periodic orbit for chaotic systems \cite{berry},
or than the inverse time of diffusion through the sample 
in disordered metallic systems \cite{shklovskii}. Another 
time scale which is absent in RMT is the Ehrenfest time $\tau_E$
\cite{berman}, i.e. the time it takes for the underlying 
classical chaotic dynamics
(with Lyapunov exponent $\lambda$) to stretch an initial narrow
wavepacket, of spatial extension given by the Fermi wavelength
$\lambda_F$, to the linear system size $L$. Defining
$M=L/\lambda_F$, one has
$\tau_E=\lambda^{-1} \ln[M/(2\tau_D)^2]$ \cite{vavilov,caveat1},
with $\tau_D = M/2 N$, the dwell time through the cavity
(all times will be measured in units of the time of flight
across the cavity). Note, in particular, 
that the growth of $\tau_E$ in the semiclassical
limit $M \rightarrow \infty$ is only logarithmic.
The emergence of a finite $\tau_E/\tau_D$ leads to strong deviations from the 
universal RMT behavior, and in particular
to the suppression of shot noise \cite{agam,silvestrov,tworzydlo}, or 
the proximity gap in Andreev 
billiards \cite{vavilov,lodder,jacquod,kormanyos}. It has 
furthermore been predicted
that weak localization vanishes at large $\tau_E/\tau_D$ \cite{aleiner}.
Also, in dirty d-wave
superconductors, the RMT behavior of the quasiparticle density of states
\cite{simons} is restored only below an energy scale set by $\tau_E$
\cite{adajac}. 

The suppression of shot-noise for $\tau_E/\tau_D \rightarrow \infty$ 
is due to the disappearance of the stochasticity of quantum mechanical
transport, and its replacement by the determinism of classical 
transport \cite{agam,silvestrov,tworzydlo}.
Wavepackets traveling on 
scattering trajectories shorter than $\tau_E$ have no time to diffract, and
are thus either fully transmitted or fully reflected. 
In an open chaotic cavity, different
scattering trajectories have in general different dwell times 
with a distribution $p(t)=\exp(-t/\tau_D)/\tau_D$. 
For finite $0 <  \tau_E/\tau_D \ll \infty $, this
suggests that transport is mediated by a two-phase dynamical fluid,
consisting of a stochastic phase of relative volume
$\alpha \simeq \int_{\tau_E}^{\infty} p(t) dt = \exp(-\tau_E/\tau_D)$, and a
deterministic phase of relative volume $1-\alpha$. Following this
purely classical argument, first expressed in Ref.\ \cite{silvestrov},
one expects that 
the distribution of transmission eigenvalues is given by
\begin{equation}\label{probtalpha}
P_{\rm \alpha}(T) = \alpha P_{\rm RMT}(T) +\frac{1-\alpha}{2}
\left[\delta(T)+\delta(1-T)\right].
\end{equation}
This will be confirmed below. We will 
further show that, quite surprisingly, the two-phase dynamical fluid 
assumption 
also correctly describes the behavior of mesoscopic coherent effects,
and in particular that it explains the breakdown of universality
of the conductance fluctuations when $\tau_E$ becomes comparable
to the ergodic time $\tau_0$.

We first summarize our main results. (i) 
We give full confirmation of Eq.~(\ref{probtalpha}) by calculating
the integrated distribution of transmission eigenvalues
$I(T) \equiv \int_0^T P(T') dT'$. We find that it is very well
fitted by (see the inset to Fig.1)
\begin{equation}\label{intprobtalpha}
I_{\alpha}(T) =\frac{1-\alpha}{2}(1+\delta_{1,T})+ 
\frac{2 \alpha}{\pi}\sin^{-1}(\sqrt{T}),
\end{equation}
from which we extract $\alpha \simeq \exp(-\tau_E/\tau_D)$. 
(ii) The conductance fluctuations stay at their universal value, 
independently on $\tau_E/\tau_D$, under variation of the energy in
a given sample. This follows from the survival of a large number
of stochastic channels -- even though their relative measure
$\alpha \rightarrow 0$ -- which preserves the universality of the conductance 
fluctuations. 
(iii) A completely different
situation arises when one considers sample-to-sample fluctuations.
In this case,
one has $\sigma^2(g) \propto (M/M_c)^2$ for $M > M_c$. The
scaling parameter  
$M_c = \tau_D^2 \exp(\lambda)$ is determined by the 
quantum mechanical resolution of classical phase space structures
corresponding to the largest cluster of fully transmitted or reflected
neighboring trajectories (see Ref.~\cite{silvestrov}).
(iv) The energy conductance correlator always decays on the
universal scale of the Thouless energy, 
$\xi_{\varepsilon} \propto 1/\tau_D$,
independently on $\tau_E$. The results (ii) and (iii) show that
the ergodic hypothesis breaks down as $\tau_E/\tau_0$ increases.
Accordingly, (iv) is somewhat surprising, but will be understood below
via a semiclassical argument.
All our results and arguments fully confirm
the two-phase dynamical fluid model. 
We note that our conclusion (iii) is in agreement with the very recent
finding $\sigma^2(g) \propto M^2$ obtained by
Tworzyd{\l}o, Tajic and Beenakker \cite{jakub}. However, (ii)
is in complete opposition with their prediction that deviations from
$\sigma^2(g)=1/8$ should occur 
upon variation of the energy for large $\tau_E/\tau_D$. Points 
(i) and (iv) are addressed here for the first time.

We consider open systems with fully developed chaotic dynamics,
for which $\tau_D \gg 1$. Because $\tau_E$ grows 
logarithmically with $M$, 
and since we want to investigate the regime $\tau_E/\tau_D \gtrsim 1$, 
we model the electron dynamics
by the kicked rotator map, which reproduces 
most of the phenomenology of low-dimensional 
noninteracting electronic physics \cite{fishman}.
The classical kicked rotator map is given by
\begin{eqnarray}\label{clkrot}
\left\{\begin{array}{lll}
\bar{x} & = & x + p \\
\bar{p} & = & p + K \sin(\bar{x}),
\end{array} \right.
\end{eqnarray}
with $K$ the (dimensionless) kicking strength. It drives the
dynamics from fully
integrable ($K=0$) to fully chaotic [$K\agt 7$, with Lyapunov exponent
$\lambda\approx\ln (K/2)$]. 
We consider a toroidal classical phase space $x,p \in[0,2 \pi]$, 
and open the system by
defining contacts to ballistic leads via two absorbing phase space strips
$[x_L-\delta x,x_L+\delta x]$ and $[x_R-\delta x,x_R+\delta x]$, 
each of them with a width $2 \delta x=\pi/\tau_D$.

Quantizing the map
amounts to a discretization of, say, the 
real space coordinates as $x_m=2 \pi m/M$, $m=1,\ldots M$. 
A quantum representation of the map (\ref{clkrot}) is
provided by the unitary $M \times M$ 
Floquet operator $U$ \cite{reichl}, 
which gives the time evolution for one iteration of the map. 
For our specific choice of the kicked rotator,
the Floquet operator has matrix elements
\begin{eqnarray}\label{kickedU}
U_{m,m'} &=& M^{-1/2} e^{-(iMK/4\pi)[\cos(2\pi m/M)+\cos(2\pi m'/M)]} 
\nonumber \\
&& \times \sum_l e^{2 \pi i l(m-m')/M} e^{-(\pi i/2M) l^{2}}.
\end{eqnarray}
The spectrum $\exp(i \varepsilon_\alpha)$ of $U$
defines a discrete set of $M$ quasienergies $\varepsilon_\alpha \in[0,2 \pi)$
with an average level spacing $\delta = 2 \pi/M$. 

Much in the same way as in the Hamiltonian case \cite{iida}, a 
quasienergy-dependent $2N \times 2N$ scattering matrix can be determined 
from the Floquet operator $U$ as \cite{fyodorov}
\begin{equation}\label{smatrix}
S(\varepsilon) = P [\exp(-i \varepsilon) - U (1-P^T P)]^{-1} U P^T,
\end{equation}
using a $2N \times M$ projection matrix $P$ which
describes the coupling to the leads. 
Its matrix elements are given by
\begin{eqnarray}\label{lead}
 P_{n,m}=\left\{\begin{array}{ll}
1& \mbox{if $n=m \in \{m_i^{(L)} \} \bigcup \; \{m_i^{(R)} \}$}\\
0& \mbox{otherwise.}
\end{array}\right.
\end{eqnarray}
An ensemble of samples with the same microscopic properties
can be defined by varying the position $\{m_i^{(L,R)} \}$, $i=1, \ldots N$ 
of the contacts to the left and right leads for fixed $\tau_D=M/2N$ and $K$.

As usual, the scattering matrix can be written in a four block form
in term of $N \times N$ transmission and reflection matrices as
\begin{eqnarray}\label{blocks}
S = \left( \begin{array}{ll}
r& t \\
t'& r' 
\end{array}\right).
\end{eqnarray}
The spectrum of transmission probabilities is given
by the $N$ eigenvalues $T_i$ of $\hat{T} = tt^{\dagger}$, from which
the dimensionless conductance is obtained, via the Landauer
formula $g = \sum_i T_i$ \cite{buttiker}. 
Our numerical procedure follows the description
given in Ref.~\cite{tworzydlo}.

We plot in Fig.1 various distributions $P(T)$ of transmission eigenvalues.
First, it is seen that 
our model correctly reproduces the RMT distribution of Eq.~(\ref{probt})
in the limit $\tau_E/\tau_D \ll 1$. The distribution undergoes strong
modifications, however, as $\tau_E/\tau_D$ increases. In particular,
more and more weight is accumulated at $T=0$ and 1. The behavior exhibited
by $P(T)$ for finite $\tau_E/\tau_D$ seems very similar to that predicted
by Eq.~(\ref{probtalpha}). To confirm this, we calculate the integrated
probability distribution $I(T)$, which presents the advantage of being
independent on the size of histogram bins.
Results are shown in the inset to Fig.~1.
The fitting curves clearly confirm
the validity of Eq.~(\ref{probtalpha}). The extracted
parameter $\alpha$ is found to obey  
$\alpha \approx \exp(-(1+\tau_E)/\tau_D)$, for $\tau_E>0$. 
We attribute the factor $1 + \tau_E$ in the exponential
(and not $\tau_E$) to the discrete nature of the dynamics in our
model. 

\begin{figure}
\includegraphics[width=5.9cm,angle=270]{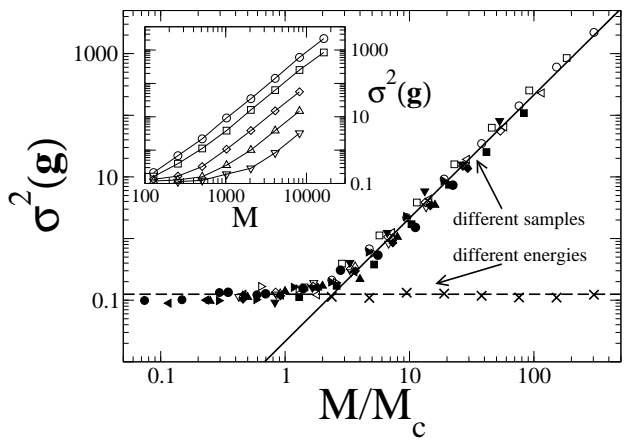}
\caption{Variance $\sigma^2(g)$ 
of the conductance vs. $M/M_c$, for
microscopic parameters $K \in [9.65,27.65]$, $\tau_D \in [5,25]$,
and $M \in [128,16384]$. The scaling parameter 
$M_c\tau_D^2 \exp(\lambda)$ varies by a factor $70$.
The solid and dashed lines indicate the classical, sample-to-sample
behavior $\propto M^2$, 
and the universal behavior  $\sigma^2(g)=1/8$ 
respectively. 
Inset: unscaled data
for $K=9.65$ and $\tau_D =5$ (circles), 7 (squares), 10 (diamonds),
15 (upward triangles) and 25 (downward triangles).}
\label{figure1}
\end{figure}

For $\tau_E \rightarrow 0$, one is in the UCF regime,
where the conductance fluctuates 
equivalently from sample to sample or as $\varepsilon$ is varied
within a given sample. This is no longer the case, however, once $\tau_E$ 
becomes finite, as is shown in Fig.~2. While $\sigma^2(g) = 1/8$ seems
to be preserved when $\varepsilon$ is varied for a given sample, one gets
an enormous increase $\sigma^2(g) \propto M^2$ from sample to sample.
This behavior derives from the underlying classical dynamics, and
can be understood on the basis of a two-phase dynamical fluid,
as we now proceed to explain.

\begin{figure}
\includegraphics[width=7.7cm,angle=0]{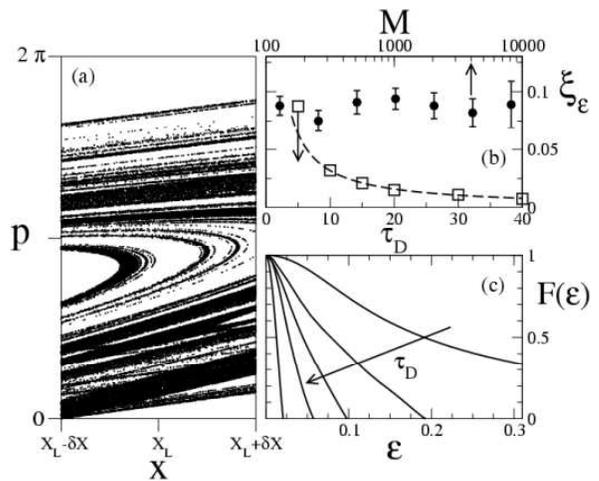}
\caption{(a) Phase space cross section of the left lead for
$K=9.65$, and $\tau_D=5$.
Black dots indicate transmitted, white areas
reflected classical trajectories respectively.
We used 25000 initial conditions, and only trajectories exiting
the system after less than 5 iterations of the classical map
(\ref{clkrot}) have been kept.
(b) Correlation length $\xi_{\varepsilon}$ extracted from the 
conductance correlator as $F(\xi_{\varepsilon})=0.8$, for
$K=9.65$ and $\tau_D=5$ vs. $M$ (black circles) and for 
$K=9.65$ and $M=2048$ vs. $\tau_D$ (white squares). 
The dashed line indicates the expected $\sim 1/\tau_D$ behavior (see text). 
(c) Decay of the conductance correlator $F(\varepsilon)$ vs. 
quasienergy for $K=9.65$, $M=2048$, and $\tau_D=5$, 10, 15, 
20, and 40.}
\label{figure2}
\end{figure}

While the classical dynamics considered here is fully chaotic,
finite-sized phase space structures emerge due to the opening of the
cavity. Following
Ref.~\cite{silvestrov}, these structures can be visualized by marking 
which trajectories originating from, say, the left lead, ends
up being transmitted to the right lead.
Such a picture is shown in Fig.~3(a). It is seen that
classical trajectories are transmitted by a series of bands exiting
the cavity at times $t_j$, and covering
a phase space area $A_j \simeq \tau_D^{-2} \exp(-\lambda t_j)$. 
The shape, position and precise volume 
of these bands is sample-to-sample dependent.
In the semiclassical limit
$\hbar_{\rm eff} = 2 \pi/M \rightarrow 0$, the effective 
Planck scale $\hbar_{\rm eff}$ resolves the
$j^{\rm th}$ band as soon as $\hbar_{\rm eff} \le A_j$, or
$M \ge 2 \pi \tau_D^{2} \exp(\lambda t_j)$. Once the largest
classical band is resolved, $\sigma^2(g)$ starts to be dominated by
the band fluctuations. Each resolved band 
carrying a growing number $\propto M$ of fully
transmitted (or reflected) quantum mechanical modes, one expects 
a variance $\sigma^2(g) \propto (M/M_c)^2$ for sufficiently large $M>M_c$,
with a scaling parameter $M_c=2 \pi \tau_D^{2} \exp(\lambda)$ determined 
by the largest band,
exiting the system at the ergodic time $\tau_0 \approx 1$. As shown on Fig.~2, 
this is precisely what happens \cite{caveat}. Deviations from 
the universal behavior emerge for $M \approx M_c$, equivalently when 
$\tau_E/\tau_0 \approx 1$, i.e. much earlier than the suppression of shot noise
(see also Ref.~\cite{jakub}).

We face a completely different situation when varying $\varepsilon$ 
within a given sample. In a ballistic system like ours, such a change 
does not modify the classical trajectories, and thus
alters only the action phase of each contribution
to the semiclassical Green function. Within this picture, the 
conductance fluctuates only due to
long, diffracting orbits with $t>\tau_E$ \cite{argaman}.
These long orbits build up the stochastic phase. 
Their subset can be viewed
as corresponding to an effective stochastic cavity with contacts to leads 
with $N_{\rm eff} = \alpha N$ channels. Fixing the microscopic parameters
$\tau_D$ and $\lambda$, one has 
$N_{\rm eff} \sim M^{1-1/\lambda \tau_D} \gg 1$. This means that,
despite the prefactor $\alpha$, $N_{\rm eff}$
is always large enough to guarantee that transport occurs
semiclassically, and therefore, one stays always in a regime with 
universal value
$\sigma^2(g)=1/8$ \cite{ucfrmt}. A first confirmation of this argument is
provided by the numerical data shown in Fig.~2, which indicate 
a constant behavior of $\sigma^2(g)$, independently 
on $\tau_E/\tau_D$. To further check this
argument, we finally consider the conductance correlator 
\begin{eqnarray}
F(\varepsilon) & = &  \sigma^{-2} (g) \; \langle \delta g(\varepsilon_0)
\delta g(\varepsilon_0+\varepsilon) \rangle .
\end{eqnarray}
As said above, only the phase accumulated 
after diffraction (for $t>\tau_E$) contributes to conductance
fluctuations \cite{argaman}, and since the subset of diffractive 
trajectories have an average dwell time given by $\tau_E+\tau_D$, they
accumulate a relevant relative phase $\propto \varepsilon \; \tau_D$.
One therefore expects a decay of $F(\varepsilon)$ over the Thouless 
scale as in the universal regime \cite{ucf}, 
$\xi_{\varepsilon} \propto 1/\tau_D$, independently
on $\tau_E$. This is confirmed by the
data shown in Fig.~3(b) and (c).

Our results thus show that the separation
of the deterministic and stochastic phases is complete. 
Beyond a simple explanation
of the suppression of shot-noise with $\alpha = \exp(-\tau_E/\tau_D)$ via 
Eq.(\ref{probtalpha}), the phase separation correctly
accounts for the behavior of the conductance variance 
and correlators in open quantum chaotic systems in the semiclassical
limit. Further investigations along the lines initiated
here should focus on other effects of mesoscopic coherence, such
as the weak-localization corrections, and in particular 
the magnetoresistance.

We thank Marlies Goorden for a clarifying discussion of 
Ref.~ \cite{silvestrov},
and Carlo Beenakker for sending us a preprint of Ref.~ \cite{jakub}.
This work was supported by the Swiss National Science Foundation.


\begin{thebibliography}{99}
\bibitem{caveat1} The deviation from the closed sample result
$\tau_E=\lambda^{-1} \ln (M)$ is due
to partial reflection at the boundary between the
cavity and the contacts to the leads \cite{vavilov}.
\bibitem{caveat} The estimate
$A_j \simeq \tau_D^{\beta} \exp(-\lambda t_j)$ with $\beta=2$
strictly applies when the beam of 
electrons injected into the cavity has a similar spread in momentum
and real-space coordinates. For the data shown in Fig.~2, we
found a significantly better numerical rescaling with
$\beta \approx 1.5$ (not shown), which we relate
to the anisotropy in the electron injection beam.
\end{thebibliography}
\end{document}